\documentclass{article}

\usepackage{times}
\usepackage{graphicx}
\usepackage{natbib}
\usepackage{amssymb}
\usepackage{amsmath}
\usepackage{dsfont}
\usepackage{amsthm}
\usepackage{algorithm}
\usepackage{algorithmic}
\usepackage{hyperref}

\usepackage[accepted]{icml2015}

\begin{document} 

\twocolumn[
\icmltitle{Bayesian and Empirical Bayesian Forests}

\icmlauthor{Matt Taddy}{taddy@chicagobooth.edu}
\icmladdress{University of Chicago Booth School of Business}
\icmlauthor{Chun-Sheng Chen}{chunschen@ebay.com}
\icmladdress{eBay}
\icmlauthor{Jun Yu}{junyu@ebay.com}
\icmladdress{eBay}
\icmlauthor{Mitch Wyle}{mwyle@ebay.com}
\icmladdress{eBay}

\icmlkeywords{decision trees, random forests, CART, 
nonparametric Bayesian, ensemble learning, 
big data, distributed computing, spark}

\vskip 0.3in
]

\begin{abstract} 
We derive ensembles of decision trees through a nonparametric Bayesian model,
allowing us to view random forests as samples from a posterior distribution.
This insight provides large gains in interpretability, and motivates a class
of Bayesian forest (BF) algorithms that yield small but reliable performance
gains. Based on the BF framework, we are able to show that high-level tree
hierarchy is stable in large samples. This leads to an empirical Bayesian
forest (EBF) algorithm for building approximate BFs on massive distributed
datasets and we show that EBFs outperform sub-sampling based alternatives by a
large margin.
\end{abstract} 

\section{Introduction}\label{introduction}

Decision trees are a fundamental machine learning tool. They partition the
feature (input) space into regions of response homogeneity, such that the
response (output) value associated with any point in a given partition can be
predicted from the average for that of its neighbors. The classification and
regression tree (CART) algorithm of \cite{breiman_classification_1984} is a
common recipe for building trees; it grows greedily through a series of
partitions on features, each of which maximizes reduction in some measure of
impurity at the current tree leaves (terminal nodes; i.e., the implied input
space partitioning). The development of random forests (RF) by
\cite{breiman_random_2001}, which predict through the average of many CART
trees fit to bootstrap data resamples, is an archetype for the
successful strategy of tree ensemble learning. For prediction problems with
training sets that are large relative to the number of inputs, properly trained ensembles of
trees can predict out-of-the-box as well as any carefully tuned,
application-specific alternative.

This article makes three contributions to understanding and application of
decision tree ensembles (or, \textit{forests}).

\textit{Bayesian forest:}  A nonparametric Bayesian (npB) point-of-view
allows interpretation of forests as a sample from a posterior over trees.
Imagine CART applied to a data generating process (DGP) with finite support:
the tree greedily splits support to minimize impurity of the partitioned
response distributions (terminating at some minimum-leaf-probability
threshold). We present a nonparametric Bayesian model for DGPs based on
multinomial draws from (large) finite support, and derive the Bayesian forest
(BF) algorithm for sampling from the distribution of CART trees implied by the
posterior over DGPs.  Random forests are an approximation to this BF exact
posterior sampler, and we show in examples that BFs provide a small but
reliable gain in predictive performance over RFs.

\textit{Posterior tree variability:} Based upon this npB framework, we derive
results on the stability of CART  over different DGP realizations. We find
that, conditional on the data allocated to a given node on the sample CART tree,
the probability that the next split for a posterior DGP realization  matches
the observed full-sample CART split is
\begin{equation}\label{treestable}
\mathrm{p}\left(\text{split matches sample CART}\right) \gtrsim 1 - \frac{p}{\sqrt{n}} e^{-n},
\end{equation}
where $p$ is the number of possible split locations and $n$ the number of observations on the  node.  Even if $p$ grows with $n$, the result indicates that partitioning can be stable conditional on the data being split.  This conditioning is key: CART's well known instability is due to its recursive nature, such that a single split different from sample CART at some node removes any expectation of similarity  below that node.  However, for  large samples, (\ref{treestable}) implies that we will see little variation at the top hierarchy of  trees in a forest.  We illustrate such stability in our examples.

\textit{Empirical Bayesian forests:} the npB forest
interpretation and tree-stability results lead us to propose empirical
Bayesian forests (EBF) as an algorithm for building approximate BFs on massive
distributed datasets.
Traditional empirical Bayesian analysis fixes parameters in high levels of a
hierarchical model at their marginal posterior mode, and quantifies
uncertainty for the rest of the model conditional upon these fixed estimates.
EBFs work the same way: we fit a single shallow CART \textit{trunk} to the
sample data, and then sample a BF ensemble of \textit{branches} at each
terminal node of this trunk.  The initial CART trunk thus maps observations to
their branch, and each branch BF is fit in parallel without any
communication with the other branches.  With little posterior
variability about the trunk structure, an EBF sample should look similar to
the (much more costly, or even infeasible) full BF sample.  In a number of
experiments, we compare EBFs to the common distributed-computing strategy of
fitting forests to data subsamples and find that the EBFs lead to a large
improvement in predictive performance.  This type of strategy is 
key to efficient machine learning with Big Data: focus the `Big' on the pieces
of models that are most difficult to learn.


Bayesian forests are introduced in Section 2 along with a survey
of  Bayesian tree models, Section 3  investigates tree stability in
theory and practice, and Section 4 presents the empirical Bayesian forest
framework. Throughout, we use publicly available data on home prices in
California to illustrate our ideas. We also provide a variety of other data
analyses to benchmark performance, and close with description of how EBF
algorithms are being built and perform in large-scale machine learning at
eBay.com.

\section{Bayesian forests}\label{bayesian-forests}

Informally, write $dgp$ to represent the stochastic process defined over a set of possible DGPs.
A Bayesian analogue to classical `distribution-free' nonparametric
statistical analysis \citep[e.g.,][]{hollander_nonparametric_1999} has two components:
\begin{enumerate}
\item set a nonparametric statistic $\mathcal{T}(dgp)$ that is of interest in your application regardless of the true DGP,
\item and build a flexible model for the DGP, so that the posterior distribution on $\mathcal{T}(dgp)$ can be derived from  posterior distribution on possible DGPs.
\end{enumerate}
In  the context of this article, $\mathcal{T}(dgp)$ refers to a CART tree.   Indeed, trees are useful precisely because they are good predictors regardless of the underlying data distribution -- they do not rely upon distributional assumptions to share information across training observations. 
Our DGP model, detailed below, leads to a posterior for $dgp$ that is represented through random weighting of observed support.  A Bayesian forest contains CART fits corresponding to each draw of support weights, and the BF ensemble prediction is an approximate posterior mean.

\begin{figure*}
\includegraphics[width=\textwidth]{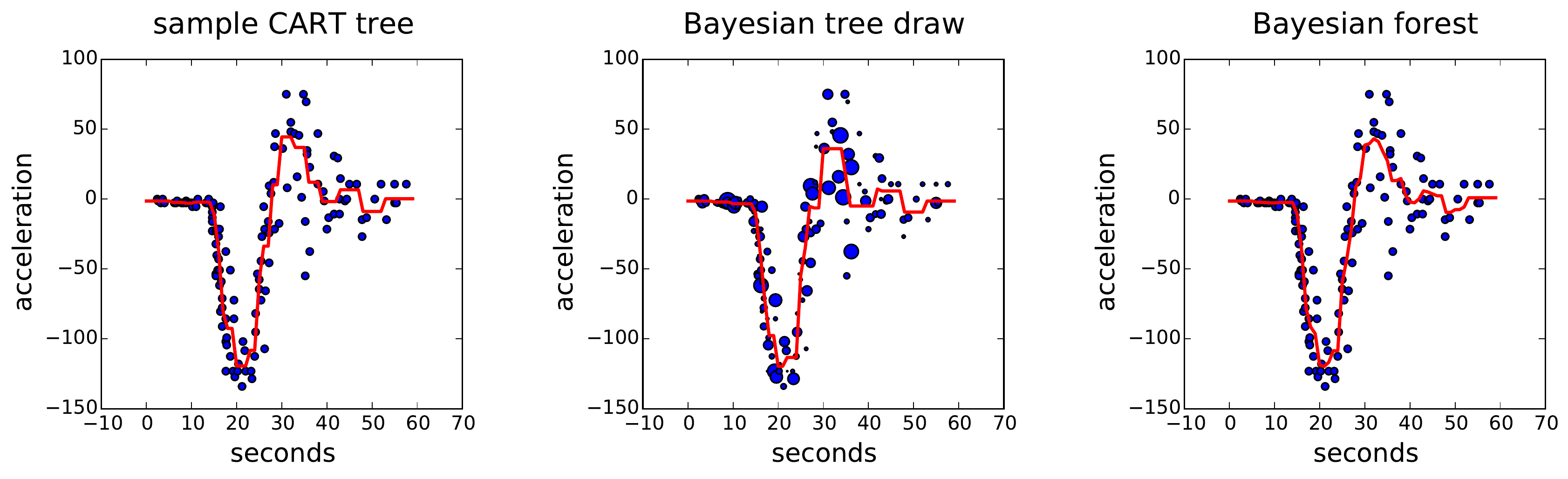}   
\caption{\label{mcycle} Motorcycle data illustration.  The lines show each of CART fit to the data sample, CART fit to a sample with weights drawn IID from an exponential (point size proportional to weight), and the BF predictor which averages over 100 weighted CART fits.} 
\end{figure*}    

\subsection{Nonparametric model for the DGP}\label{dgpmodel}

We employ a Dirichlet-multinomial sampling model in nonparametric
Bayesian analysis. The approach dates back to
\citet{ferguson_bayesian_1973}. \citet{chamberlain_nonparametric_2003}
provide an overview in the context of econometric problems.
\citet{rubin_bayesian_1981} proposed the Bayesian bootstrap as an
algorithm for sampling from versions of the implied posterior, and it has since become closely associated with
this model.

Use $\mathbf{z}_i = \{\mathbf{x}_i,y_i\}$ to denote the  features
and response  for observation $i$. We suppose that data are drawn
\emph{independently} from a finite  $L$  possible values,
\begin{equation}\label{dgpeq}
dgp = \mathrm{p}(\mathbf{z}) = \sum_{l=1}^L \omega_l \mathds{1}_{[\mathbf{z} = \boldsymbol{\zeta}_l]}
\end{equation}
where $\omega_l\geq0\forall l$ and $\sum_l \omega_l = 1$. Thus the generating
process for observation $i$ draws $l_i$ from a multinomial with probability
$\omega_{l_i}$, and this indexes one of the $L$ support points. Since $L$ can
be arbitrarily large, this so-far implies no restrictive assumptions beyond
that of independence.

The conjugate prior for $\boldsymbol{\omega}$ is a Dirichlet
distribution, written
$\mathrm{Dir}(\boldsymbol{\omega}; \boldsymbol{\nu}) \propto \prod_{l=1}^L\omega_j^{\nu_l-1}$.
We will parametrize the prior with a single concentration parameter
$\boldsymbol{\nu} = a >0$, such that $\mathbb{E}[\omega_l] = a/La = 1/L$
and $\mathrm{var}(\omega_l) = (L-1)/[L^2(La+1)]$. Suppose you have the
observed sample $\mathbf{Z} = [\mathbf{z}_1 \cdots \mathbf{z}_n]'$. For
convenience, we allow $\boldsymbol{\zeta}_l=\boldsymbol{\zeta}_k$ for
$l \neq k$ in the case of repeated values. Write
$l_1 \dots l_n = 1 \dots n$ so that
$\mathbf{z}_i = \boldsymbol{\zeta}_i$ and
$\mathbf{Z} = [\boldsymbol{\zeta}_1 \cdots \boldsymbol{\zeta}_n]'$. Then
the posterior distribution for $\boldsymbol{\omega}$ has
$\omega_i = a+1$ for $i\leq n$ and $\omega_l = a$ for $l>n$, so that
\begin{equation}\label{omegapost}
\mathrm{p}(\boldsymbol{\omega}) \propto \prod_{i=1}^n \omega_i^{a} \prod_{l=n+1}^L \omega_l^{a-1}.
\end{equation}
This, in turn, defines our posterior for the data generating process
through our sampling model in (\ref{dgpeq}).

There are many possible strategies for specification of $a$ and
$\zeta_l$ for $l>n$.\footnote{The unobserved $\zeta_l$ act as data we imagine we
might have seen, to smooth the posterior away from the data we have
actually observed. See \citet{poirier_bayesian_2011} for discussion.} The non-informative prior
that arises as $a\rightarrow 0$ is a convenient default: in this limit, $\omega_l = 0$ with probability one for
$l>n$.\footnote{For $l>n$ the posterior has
$\mathbb{E}[\omega_l]=0$ with variance
$\mathrm{var}(\omega_l) = \lim_{a \to 0} a[n+a(L-1)]/[(n+La)^2(n+La+1)] = 0$.}
We apply this limiting prior throughout, such that our posterior for the
data generating process is a multinomial draw from the observed
data points, with a uniform $\mathrm{Dir}(\boldsymbol{1})$ distribution
on the $\boldsymbol{\omega} = [\omega_1 \dots \omega_n]'$ sampling
probabilities.  We will also find it convenient to parametrize un-normalized $\boldsymbol{\omega}$ via
IID exponential random variables: $\boldsymbol{\theta} = [\theta_1 \dots \theta_n]$, where $\theta_i \stackrel{ind}{\sim} \mathrm{Exp}(1)$ in the posterior and $\omega_i = \theta_i/|\boldsymbol{\theta}|$ with $|\boldsymbol{\theta}| = \sum_i \theta_i$.

\subsection{CART as posterior functional}

Conditional upon $\boldsymbol{\theta}$, the population tree
$\mathcal{T}(dgp)$ is defined through a weighted-sample
CART fit. In particular, given data
$\mathbf{Z}^\eta = \{\mathbf{X}^\eta,\mathbf{y}^\eta\}$ in node $\eta$,
sort through all dimensions of all observations in $\mathbf{Z}^\eta$ to
find the split that minimizes the average of some $\boldsymbol{\omega}$-weighted
impurity metric across the two new child nodes. For example, in the case
of regression trees, the impurity to minimize is
weighted-squared error
\begin{equation}
\mathcal{I}(\mathbf{y}^\eta) = \sum_{i\in \eta} \theta_i (y_i - \mu_\eta )^2
\end{equation}
with $\mu_\eta = \sum_i \theta_i y_i/|\boldsymbol{\theta}^\eta|$.   For our
classification trees, we minimize Gini impurity.  The split candidates are
restricted to satisfy a minimum leaf probability, such that every node in the
tree must have  $|\boldsymbol{\theta}^\eta|$ greater than some
threshold.\footnote{In practice this can be replaced with a threshold on the
minimum number of observations at each leaf.}   This  procedure is repeated on
every currently-terminal tree node until it is no longer possible to split
while satisfying the minimum probability threshold. To simplify notation, we
refer to the resulting CART tree as $\mathcal{T}(\boldsymbol{\theta})$.

\subsection{Posterior sampling}\label{posterior-sampling}

Following \cite{rubin_bayesian_1981} we can  sample from the
posterior on $\mathcal{T}(\boldsymbol{\theta})$ via the 
Bayesian bootstrap in Algorithm 1. 
\begin{algorithm}[ht]
   \caption{Bayesian Forest}
   \label{alg:bf}
\begin{algorithmic}
   \FOR{$b=1$ {\bfseries to} $B$}
   \STATE draw $\boldsymbol{\theta}^b \stackrel{iid}{\sim} \mathrm{Exp}(\mathbf{1})$
   \STATE run weighted-sample CART to get $\mathcal{T}_b = \mathcal{T}(\boldsymbol{\theta}^b)$
   \ENDFOR
\end{algorithmic}
\end{algorithm}

   We've implemented  BF through simple adjustment of the
\texttt{ensemble} module of python's \texttt{scikit-learn} \citep{scikit-learn}.\footnote{Replace  variable \texttt{sample\_counts} in \texttt{forest.py}  to be
drawn exponential rather than multinomial when bootstrapping.}  As a quick illustration, Figure \ref{mcycle} shows three fits for the conditional mean  velocity of a motorcyle helmet after  impact in a crash: sample CART $\mathcal{T}(\mathbf{1})$, a single draw of $\mathcal{T}(\boldsymbol{\theta})$, and the BF average prediction \citep[data are from the MASS R package,][]{mass}.

Note that the Bayesian forest differs from Breiman's random
forest only in that the weights are drawn from an exponential (or Dirichlet, when normalized) distribution
rather than a Poisson (or multinomial) distribution. To the extent that RF sampling
provides a coarse approximation to the BF samples, the former is a
convenient approximation.  Moreover, we will find little difference in predictive performance between BFs and RFs, so that one should feel free to use readily available RF software while still relying on the ideas of this paper for intuition and interpretation.

    \subsection{Bayesian tree-as-parameter
models}\label{bayesian-tree-as-parameter-models}

Other Bayesian analyses of DTs treat the tree as
a parameter which governs the DGP, rather than a functional thereof, and thus
place some set of restrictions on the distributional relationship between
inputs and outputs.   

The original Bayesian tree model is the Bayesian CART (BCART) of
\citet{chipman_bayesian_1998}.
BCART defines a likelihood where response values for  observations  allocated
(via $\mathbf{x}$) to the same leaf node are IID from some parametric family
(e.g., for regression trees, observations in each leaf are IID Gaussian with
shared variance and mean).  The  BCART authors also propose linear regression
leaves, while the Treed Gaussian Processes  of
\citet{gramacy_bayesian_2008} use Gaussian process regression at each leaf.
The models are fit via Markov chain Monte Carlo (above examples) or sequential Monte Carlo \citep{taddy_dynamic_2011} algorithms that draw from the posterior by
proposing small changes to the tree (e.g.,~grow or prune).\footnote{The Bayesian bootstrap
is also a potential sampling tool in this tree-as-parameter setting. See
\citet{clyde_bagging_2001} for details on the technique and its relation to
model averaging.}

Monte Carlo tree sampling use the same incremental moves that employed in
 CART. Unfortunately, this means that they tend to get stuck in
 locally-optimal regions of tree-space.  Bayesian Additive Regression Trees
\citep[BART;][]{chipman_bart:_2010}  replace a single tree with the sum of
many small trees. An input vector is allocated to a leaf in each  tree, and
the corresponding response distribution has mean equal to the average of each
leaf node value and  variance shared across
observations. Original BART has Gaussian errors, and extensions include
Gaussian mixtures. Since BART only  samples  short trees, it is fast
and mixes well.

Easy
sampling comes at a potentially steep price: the assumption of homoskedastic
additive errors.\footnote{For classification, this is manifest  through a
probit link.} Despite this restriction, empirical studies have repeatedly shown BART to outperform
alternative flexible prediction rules. Many response variables 
(especially after, say, log transformation) have the property that they are well fit by
 flexible regression with homoskedastic errors.  Whenever the model
 assumptions in BART are close enough to true, it should outperform  methods which do not make those assumptions.  

In contrast, the npB interpretation of
forests (BF or RF)  makes it clear that they are suited to applications where
the response distribution defies parametric representation, such that CART fit
is the most useful DGP summary available. We  often encounter this situation in
application.  For example,  internet transaction data  combines discrete
point masses  with  an incredibly fat right tail \citep[e.g.,
see][]{taddy_heterogeneous_2014}.  In academia it is common to transform such
data before analysis, but businesses wish to target the response on the scale
measured (e.g., clicks or dollars) and  need to build a predictor that does
well on that scale.

\begin{figure}
~\includegraphics[width=0.45\textwidth]{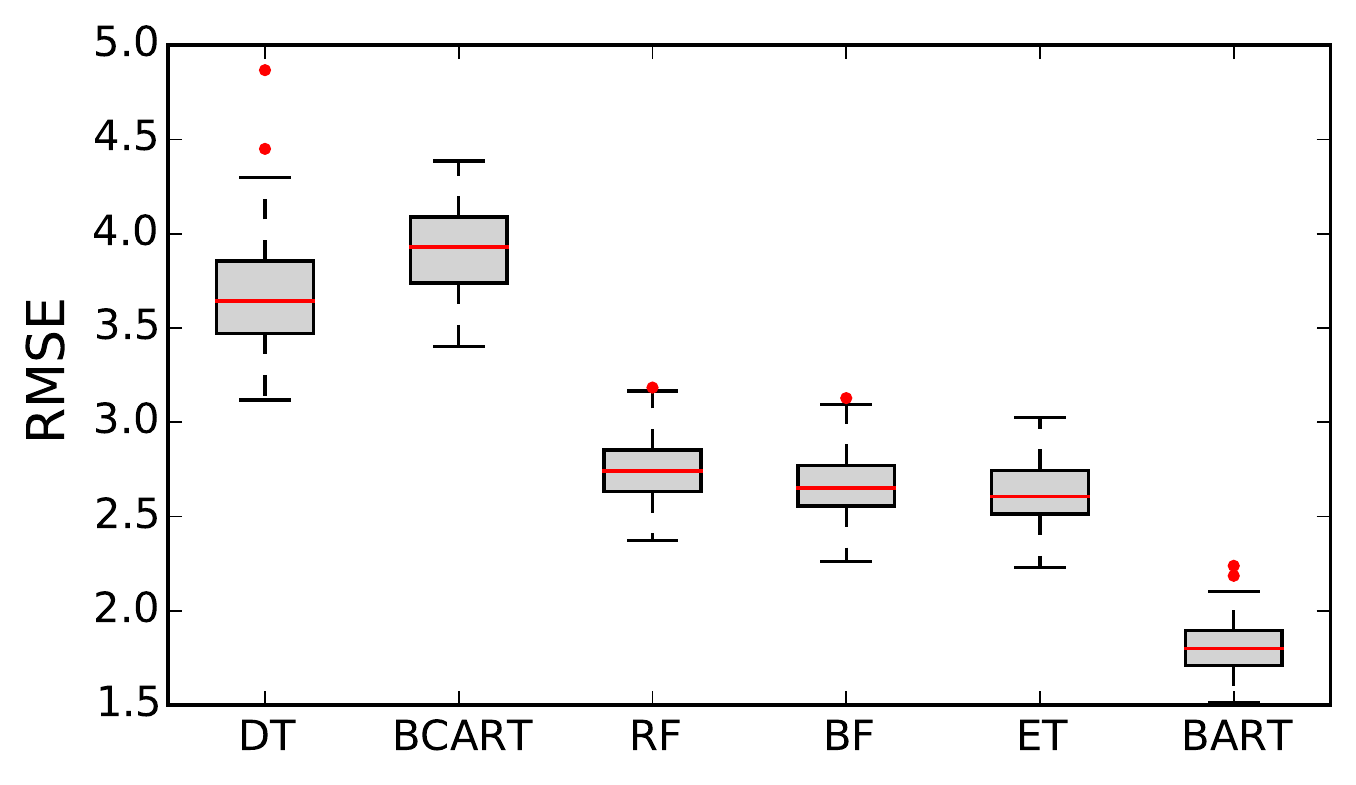}
\vskip -.25cm
\caption{\label{friedpic}Friedman experiment predictive RMSE over 100 runs.}
\end{figure}

\subsection{Friedman example}\label{friedman-example}

\begin{figure*}
~~~\includegraphics[width=0.325\textwidth]{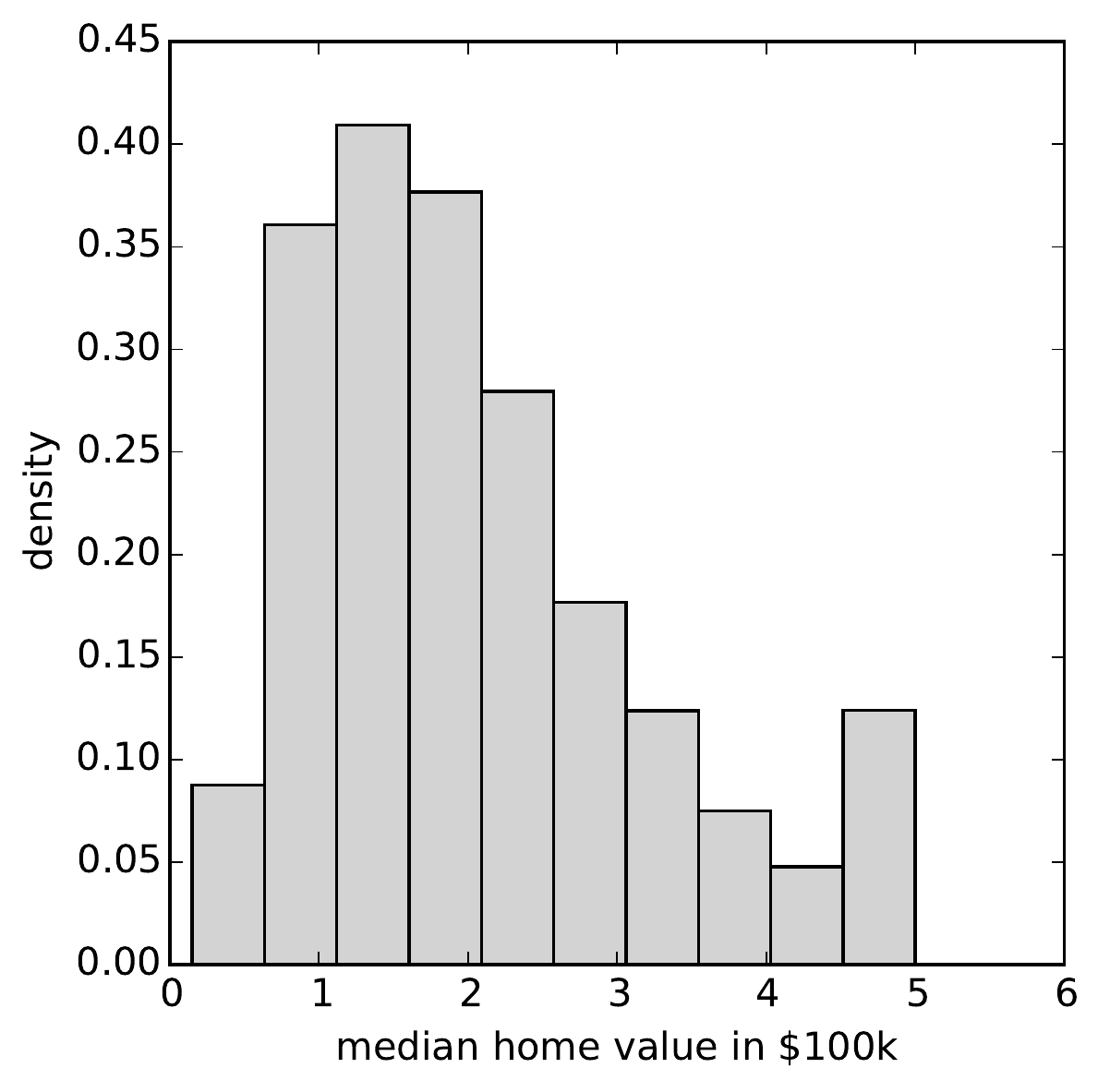}
~~~~~~~
\includegraphics[width=0.6\textwidth]{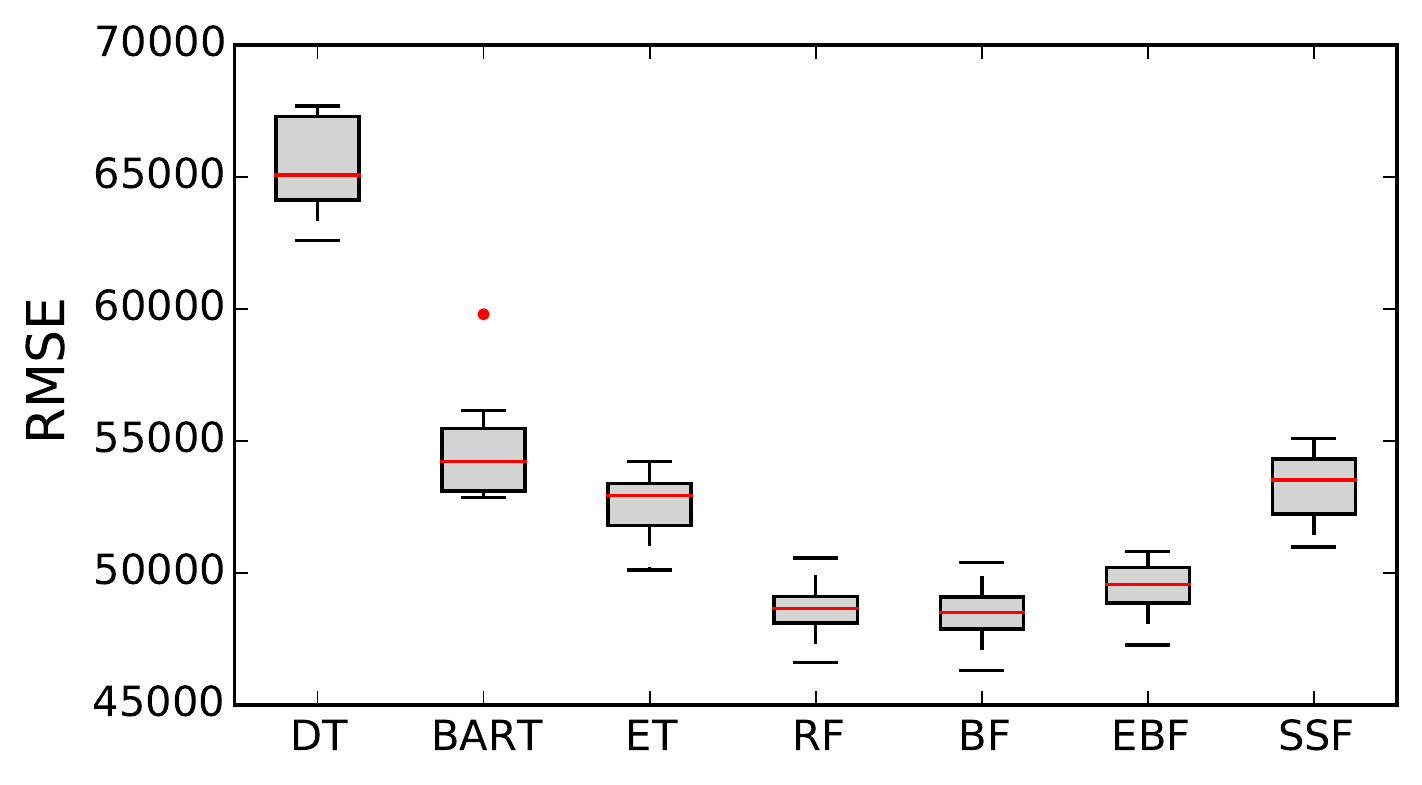}
\caption{\label{calihist} California housing data. The response sample is on the left, and the right panel shows predictive RMSE across 10-fold CV.}
\end{figure*}
    
A common simulation experiment in evaluating flexible predictors is
based around the \citet{friedman_multivariate_1991} function,
\begin{align}\label{friedfun}
y &= f(\mathbf{x}) +  \varepsilon \\ &= 10\mathrm{sin}(\pi x_1 x_2) + 20 (x_3-0.5)^2 + 10x_4 + 5x_5 + \varepsilon.\notag
\end{align}
where $\varepsilon \sim \mathrm{N}(0,1)$ and $x_j \sim \mathrm{U}(0,1)$.

For our experiment, we follow previous authors by including as features for
training the spurious $x_6 \dots x_{p}$. Each regression models are fit to 100
random draws from (\ref{friedfun}) and tested at 1000 new $\mathbf{x}$
locations. Root mean square error (RMSE) is
calculated between predicted and true $f(\mathbf{x})$.

Results over 100 repeats are shown in  Figure \ref{friedpic}.\footnote{In this and the next example, CART-based algorithms had minimum-leaf-samples set at 3 and the ensembles contain 100 trees.  BART and BCART run at their \texttt{bayestree} and \texttt{tgp} R package defaults, except that BART draws only 200 trees after a burn-in of 100 MCMC iterations.  This is done to get comparable compute times; for these settings BART requires around 75 seconds per fold with the California housing data, compared to BF's 10 seconds when run in serial, and 3 seconds running on 4 cores.}
As forecast, the only model which assumes the true homoskedastic error
structure, BART, well outperforms all others. The two forests, BF and
RF, are both a large improvement over DT. 
The BF average RMSE is only about 1\% better than the RF's,
since Bayesian and classical bootstrapping differ little in
practice. BCART does very poorly: worse than a single DT. We
hypothesis that this is due to the notoriously poor mixing of the BCART
MCMC, such that this fit is neither finding a posterior mean (as it is
intended to) or optimizing to a local posterior mode (as DT does).

We also include the extremely randomized trees (ET) of
\citet{geurts_extremely_2006}, which are similar to RFs except
that (a) instead of optimizing greedy splits, a few candidate splits are chosen
randomly and the best is used and (b) the full unweighted data sample is used to
fit each tree. ET slightly outperforms both BF and RF; in our experience this
happens in small datasets where the restriction of population support to
observed support (as assumed in our npB analysis) is invalid and the
forest posteriors are over-fit.

\subsection{California housing data}\label{california-housing-data}

As more realistic example, we consider the California housing data of \citet{calidata}
consisting of median home price along with eight features (location, income, housing stock) for 20640  census blocks. Since prices tend to vary with covariates on a multiplicative scale, standard analyses take  log median price as the response of
interest.  Instead, we will model the conditional expectation of
dollar median price.  This is relevant to applications where prediction of raw transaction data is of primary interest.  The marginal response distribution for median home price is shown in the left panel of  Figure \ref{calihist}.\footnote{Values appear to have been capped at \$500k}

Figure \ref{calihist} shows results from a 10-fold cross-validation (CV)
experiment, with details in Table \ref{calitab}.  
Except for DT and BCART, which still perform worse than all 
others\footnote{We've left off BCART's average RMSE of \$82k, 70\% WTB.},
 results are reversed from those for the small-sample and homoskedastic Friedman data.
Both RF and BF do much better than BART and it's
restrictive error model.   BF again offers a small
gain over RF.   Since the larger sample size makes observed support a better approximation to population support,  the forests outperform ET.  The EBF (empirical Bayesian forest) and SSF (sub-sample forest) predictors are based on distributed computing algorithms that we introduce later in Section 3.  At this point we note only that the massively scalable EBF is amongst the top performers; the next section helps explain why.

\begin{table}
{\footnotesize
\begin{tabular}{r|c c c c c c c}
&BF  &  RF &   EBF &  ET &   SSF &  BART & DT 
\\ \cline{2-8}\rule{0pt}{3ex}
RMSE &  48.2 &   48.5 &   49.4 &   52.5 &   53.1 &   54.8 &   65.6   
\\ \%WTB & 0.0& 0.5& 2.4& 8.7&10.0&13.4&35.9
\end{tabular}}
\caption{\label{calitab} Average RMSE in \$10k and \% worse-than-best 
 for the California housing data 10-fold CV experiment.}
\end{table}

\section{Understanding the posterior over trees}\label{treeuncertainty}

Theory on decision trees is sparse. The original CART book of
\citet{breiman_classification_1984} provides consistency results;
they show that any partitioner that is able to learn enough to
split the DGP support into very low probability leaves will be able to
reproduce the conditional response distribution of that DGP. However, this
result says nothing about the structure of the underlying trees, nor does it
say anything about the ability of a tree to predict when there is not enough
data to finely partition the DGP.  Others have focused on the frequentist
properties of individual split decisions. In the  CHAID work of
\citet{kass_exploratory_1980}, splits are based upon $\chi^2$ tests at each
leaf node. \citet{loh_regression_2002} and
\citet{hothorn_unbiased_2006} are generalizations, both of which
combat multiple testing and other biases inherent in tree-building through a
sequence of hypothesis tests. However, such contributions provide little
intuition  in the setting where we are not working from a no-split null
hypothesis distribution.

Despite this lack of theory, it is generally thought that there is large
  uncertainty (sampling or posterior) about the structure of decision trees.
  For example, \citet{geurts_investigation_2000} present extensive simulation
  of tree uncertainty and find that the location and order of splits
  is often no-better than random (this motivates work by the same
  authors on extremely randomized trees). The intuition behind such randomness
  is clear: the probability of a tree having any given branch structure is the
  product of conditional probabilities for each successive split. After 
  enough steps, any specific tree approaches probability of zero.

However, it is possible that elements of the tree structure are
 stable. For example, in the context of boosting,
\citet{appel_quickly_2013} argue that the conditionally optimal
split locations for internal nodes can be learned from
subsets of the full data allocated to each node.  They use this to propose
a faster boosting algorithm. In this section we make a related claim: 
in large samples, there is little posterior variation for the top of the tree. 
We make the point first in theory, then through empirical demonstration.

\subsection{Probability of the sample CART
tree}\label{probability-of-the-sample-cart-tree}

We focus on regression trees for this derivation, wherein node impurity is
measured as the sum of squared errors. Consider a simplified setup
with each $x_j \in \{0,1\}$  a binary random variable (possibly
created through discretization of a continuous input).  We'll investigate here the probability that the impurity minimizing split on a given node is the same for a given realization of posterior DGP weights as it is for the unweighted data sample.   

Suppose  $\{\mathbf{z}_1 \ldots \mathbf{z}_n\}$ are the data to be partitioned at some tree node, with $\mathbf{z}_i = [y_i, x_{i1}, \dots, x_{ip}]'$.  Say that
$\texttt{f}_j=\{i:x_{ij}=0\}$ and $\texttt{t}_j=\{i:x_{ij}=1\}$ are the
partitions implied by splitting on a given $x_j$. The resulting impurity is
\begin{equation}
\sigma^2_j(\boldsymbol{\theta}) = \frac{1}{n}\sum_i \theta_i \left[y_i - \mu_j(x_{ij})\right]^2,
\end{equation}
$\mu_j(0) = \sum_{i \in \texttt{f}_j}y_i \,\theta_i/\left|\boldsymbol{\theta}_{\texttt{f}_j}\right|$,
$\mu_j(1) = \sum_{i \in \texttt{t}_j}y_i \,\theta_i/\left|\boldsymbol{\theta}_{\texttt{t}_j}\right|$.

We could use the Bayesian bootstrap to simulate the posterior for $\sigma_j$ implied by the exponential
posterior on $\boldsymbol{\theta}$, but an  analytic
expression is not available. Instead, we follow the approach used in
\citet{lancaster_note_2003}, \citet{poirier_bayesian_2011}, and
\citet{taddy_heterogeneous_2014}:  derive a first-order Taylor
approximation to the function $\sigma_j$ and describe the posterior for
that related functional. 

In particular, the $1\times n$ gradient of
$\sigma_j$ with respect to $\boldsymbol{\theta}$ is
\begin{align}
\nabla \sigma^2_j & = \nabla  \frac{1}{n}\left[\sum_i \theta_iy_i^2 - \frac{1}{\left|\boldsymbol{\theta}_{\texttt{f}}\right|}\left(\mathbf{y}_{\texttt{f}}'\boldsymbol{\theta}_{\texttt{f}}\right)^2
- \frac{1}{\left|\boldsymbol{\theta}_{\texttt{t}}\right|}\left(\mathbf{y}_{\texttt{t}}'\boldsymbol{\theta}_{\texttt{t}}\right)^2\right]
\end{align}
which has elements $ \nabla_{i} \sigma^2_j = \left(y_i -
\mu(x_{ij})\right)^2/n $

The Taylor approximation is then 
\begin{align}
\sigma^2_j \approx \tilde\sigma^2_j  &= \sigma^2_j(\boldsymbol{1}) + \nabla \sigma^2\big |_{\boldsymbol{\theta}=\mathbf{1}} (\boldsymbol{\theta} - \boldsymbol{1}) \notag\\ &=  \frac{1}{n}\sum_i \theta_i \left(y_i - \bar{y}_j(x_{ij})\right)^2
\end{align} with
$\bar y_j(0) = \frac{1}{n_{\texttt{f}_j}}\sum_{i \in \texttt{f}_j}y_i$
and
$\bar y_j(1) = \frac{1}{n_{\texttt{t}_j}}\sum_{i \in \texttt{t}_j}y_i$
 the observed response averages in each partition.

    Suppose that
$\sigma_1(\boldsymbol{1}) < \sigma_j(\boldsymbol{1})~\forall j$, so that
variable `1' is that selected for splitting based on the unweighted data sample. Then we can
quantify variability about this selection by looking at differences
in approximate impurity,
\begin{align}
\Delta_j(\boldsymbol{\theta})  & = \tilde\sigma_1^2 - \tilde\sigma_j^2 
\\ & =    \frac{1}{n}\sum_i \theta_i
\big[
\bar{y}_1^2(x_{i1})-\bar{y}^2_j(x_{ij}) - \notag\\ &~~~~~~~~~~~~~~~~~~~~~~~~~~~~~~~
2y_i\left(\bar{y}_1(x_{i1})-\bar{y}_j(x_{ij})\right)\big] \notag
\\ & \equiv \frac{1}{n}\sum_i \theta_i d_{ji}. \notag
\end{align}

Say $\mathbf{d}_j = [d_{j1}\ldots d_{jn}]'$ is the vector of squared
error differences. Then the total difference has mean
$\mathbb{E}\Delta_j  = \bar{d}_j$ and variance
$\mathrm{var}\Delta_j =  \mathbf{d}_j'\mathbf{d}_j/n^2$. Since
$\Delta_j$ is the mean of independent Exponential random variables with known
means and variances, the central limit theorem applies and it
converges in distribution to a Gaussian:
\begin{equation}
\sqrt{n}(\Delta_j(\boldsymbol{\theta}) -\bar{d}_j)\rightsquigarrow \mathrm{N}(0,~\mathbf{d}_j'\mathbf{d}_j/n ).
\end{equation}
The weighted-sample impurity-minimizing split matches that for the unweighted-sample if and only if all $\Delta_j$ are negative, which
occurs with probability (note $\bar d_j < 0$)
\begin{align}
\mathrm{p}\left(\Delta_j< 0 :~ j  = 2\dots p \right) 
& \geq 1 - \sum_{j=2}^p \mathrm{p}( \Delta_j > 0 ) \\
& \rightsquigarrow 1 
 - \sum_{j=2}^p \Phi\left( -\frac{\sqrt{n}\left|\bar{d}_j\right|}{\sqrt{\mathbf{d}_j'\mathbf{d}_j/n}}\right)\notag \\ 
& \geq 1 -  \frac{1}{\sqrt{2\pi}} \sum_{j=2}^p \frac{\exp(-n z^2_j/2)}{z_j\sqrt{n}} \notag
\end{align}
where
$z_j = \left|\bar{d}_j\right|\left(\mathbf{d}_j'\mathbf{d}_j/n\right)^{-\frac{1}{2}}$
is sample mean increase in impurity over the sample standard deviation
of impurity. This ratio is bounded in probability, so that that the
exponential bound goes to zero very quickly with $n$. In particular,
ignoring variation in $z_j$ across variables, we arrive at the approximate lower bound on the probability of the sample split
\[
\mathrm{p}\left(\text{posterior split matches sample split}\right) \gtrsim 1 - \frac{p}{\sqrt{n}} e^{-n}.
\]
Even allowing for $p \approx n\times d$, with
$d$ some underlying continuous variable dimension and $p$ the input
dimension discretization on these variables, the probability of the
observed split goes to one at order $\mathcal{O}(n^{-1})$ if
$d < e^{n}/n^{3/2}$.

Given this, why is there any uncertainty at all about trees? The answer
is recursion: each split is conditionally stable given the sample at the
current node, but the probability of any specific sequence of splits is
roughly (we don't actually have independence) the product of individual
node split probabilities. This will get small as we move deeper down the
tree, and given one split different from the sample CART the
rest of the tree will grow arbitrarily far from this modal structure.
In addition, the sample size going into our probability bounds is
shrinking exponentially with each partition, whereas the dimension of
eligible split variables is reduced only by one at each level.  

Regardless of overall tree variability, we can take from this section an expectation that for large samples the
high-level tree structure varies little across posterior DGP realizations.
The next section shows this to be the case in practice.

\begin{figure*}
~~~
\includegraphics[width=0.6\textwidth]{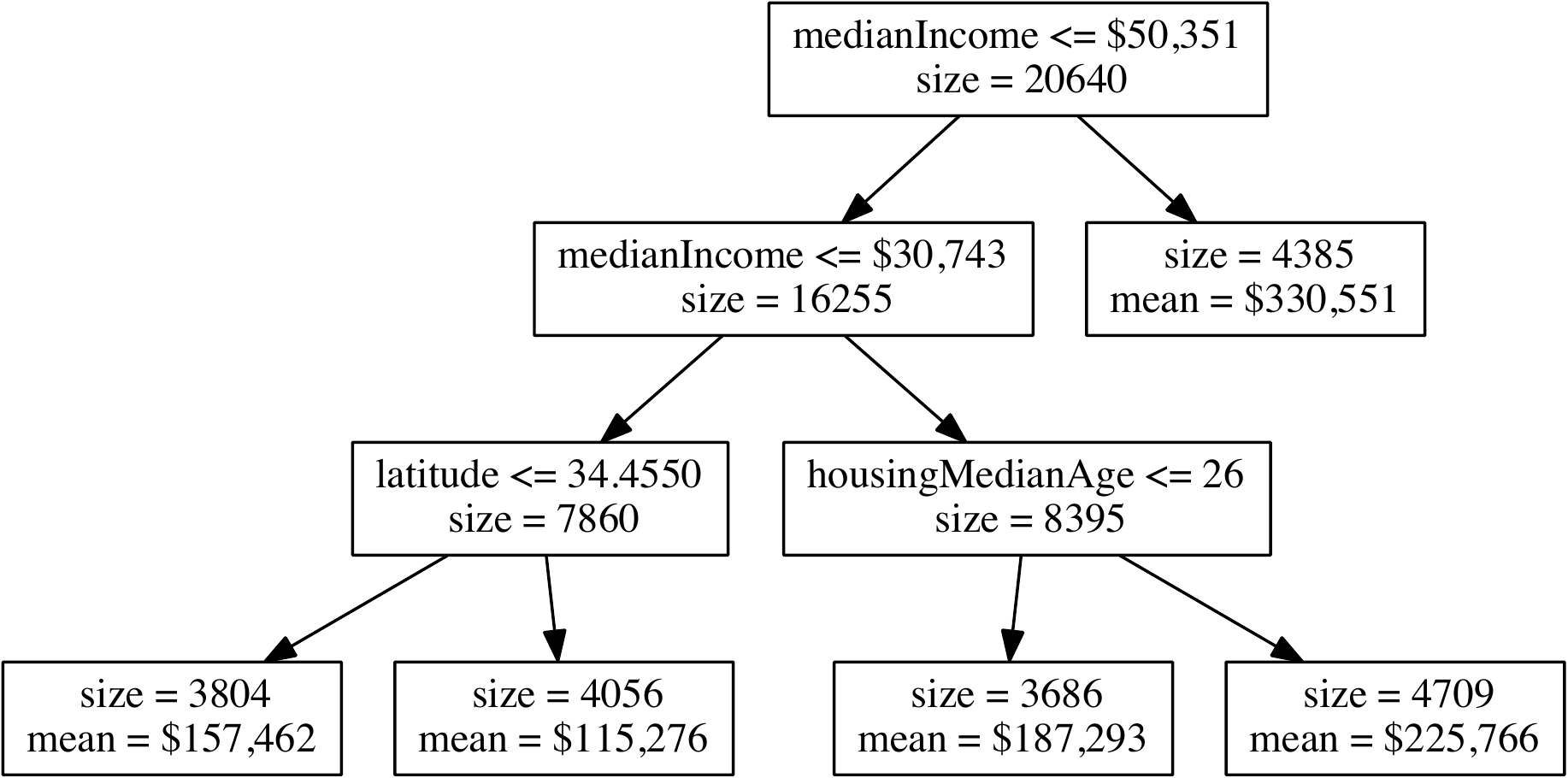}
~~~~~~
\includegraphics[width=0.31\textwidth]{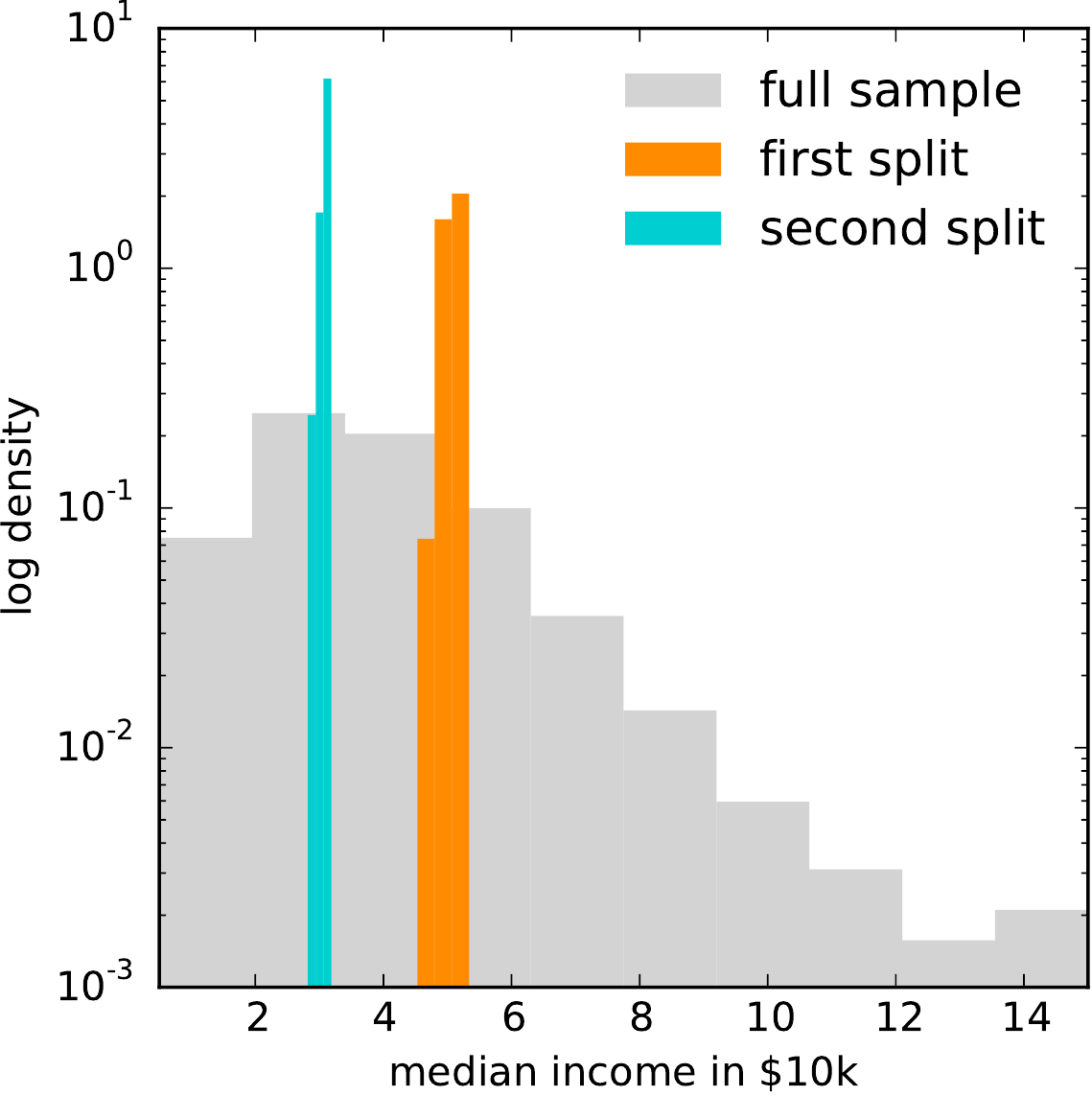}
\caption{\label{calistable} California housing data.  The left shows the CART fit  on unweighted data with a minimum of 3500 samples-per-leaf. The right panel shows the distribution of first and second split locations -- always on median income  -- across 100 draws from the BF posterior. }
\end{figure*}

\subsection{Trunk stability in California}\label{stability-in-california}

We'll illustrate the stability of high-level tree structure on our California
housing data. Consider a `trunk' DT fit to the unweighted data with no less
than 3500 census blocks (out of 20640 total) in each leaf partition.  This
leads to the tree on the left of Figure \ref{calistable} with five terminal
nodes. It splits on the obvious variable of median income, and also manages to
divide California into its north and south regions (34.455 degrees north is
just north of Santa Barbara).

To investigate tree uncertainty, we can apply the BF algorithm and repeatedly fit similar CART trees to randomly-weighted data.  Running a 100 tree BF, we find that the sample tree occurs 62\% of the time.  The second most common tree, occurring 28\% of the time, differs only in that it splits on median income again instead of on housing median age.  Thus 90\% of the posterior weight is on trees that split on income twice, and then latitude.  Moreover, a striking 100\% of trees have first two splits on median income.  From this, we can produce the plot in the right panel of Figure \ref{calistable} showing the locations of these first two splits: each split-location posterior is tightly concentrated around the corresponding sample CART split.


\section{Empirical Bayesian forests}

Empirical Bayes (EB) is an established framework with a
successful track record in fast approximate Bayesian inference; see,
e.g., \citet{efron_large-scale_2010} for an overview. In parametric
models, EB can be interpreted as fixing at their marginal posterior
maximum  the parameters at higher levels of a hierarchical
model. For example, in the simple setting of many group means (the
average student test score for each of many schools) shrunk to a global center (the outcome for an imaginary `average school'),
an EB procedure will shrink each group mean toward the overall sample average (each school towards all-student average). \citet{kass_approximate_1989}
investigate such procedures and show that, under fairly general
conditions, the EB conditional posterior for each group mean  quickly
approaches the fully Bayesian unconditional posterior as the sample size
grows.  \textit{This occurs because there is little uncertainty about the global mean.}

CART trees are not a parametric model, but they are hierarchical and
admit an interpretation similar to those studied in
\citet{kass_approximate_1989}. The data which reaches any given interior
node is a function of the partitioning implied by nodes shallower in the
tree. Moreover, due to the greedy algorithm through which CART
grows, a given shallow trunk is unaffected by changes to the tree
structure below. Finally, Section 3 demonstrated that, like high levels in a parametric hierarchical model, there is relatively little uncertainty about high levels of the tree.

An empirical Bayesian forest (EBF) takes advantage of this structure by fixing
the highest levels in the hierarchy -- the earliest CART splits -- at a single
estimate of high posterior probability.\footnote{These earliest splits -- requiring impurity search over large observation sets -- are also the most expensive to optimzie.} We fit a single
shallow CART \textit{trunk} to the unweighted sample data, and  sample a
BF ensemble of \textit{branches} at each terminal node of this trunk.  The
initial CART trunk  maps observations to their branch, and each branch BF
deals with a dataset of manageable size and can be fit in parallel without any
communication with the other branches. That is, EBF replaces the BF posterior
sample of trees with a conditional posterior sample that takes the pre-fit
trunk as fixed. Since the trunk has relatively low variance for large samples
(precisely the setting where such distribution is desirable), the EBF should
provide predictions similar to that of the full BF at a fraction of the cost.

In contrast, consider the `sub-sample forest' (SSF) algorithm, which replaces the full data with random sub-samples of manageable size.  Independent forests are  fit to each sub-sample and predictions are averaged across all forests.
SSF is a commonly applied strategy \citep[e.g., see mention, but not recommendation, of it in][]{panda_planet:_2009}, but it implies using only partial data for learning deep tree structure.  Although the tree trunks are stable, the full tree is highly uncertain and learning such structure is precisely where you want to use a full Big Data sample.

\subsection{Out-of-sample experiments}\label{oos-experiment}

In the California housing  experiment of Figure \ref{calihist}
and Table \ref{calitab}, EBF\footnote{EBFs use five node trunks  in this Section.  The SSFs are fit on data split into five equally sized subsets.} predictions are only 2\% worse than those from the full BF.  In contrast, SSF predictions are 10\% worse.

\begin{figure}
\begin{minipage}{0.5\linewidth}
\includegraphics[width=\textwidth]{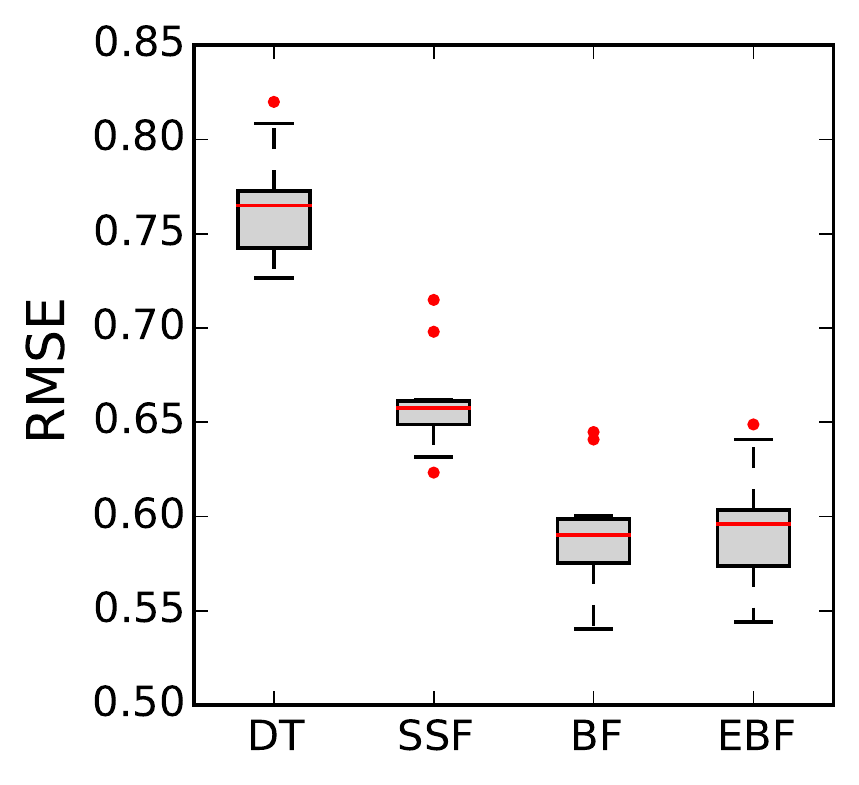}
\end{minipage}
~~
\begin{minipage}{0.4\linewidth}
{\footnotesize
\begin{tabular}{c c | l}
$\overline{\text{RMSE}}$  & \% WTB & \\
\cline{1-2}\rule{0pt}{3ex} 
0.5905 &  0.0 & BF \\
0.5953 &  0.8 & EBF \\
0.6607 & 11.9 & SSF \\
0.7648 & 29.5 & DT \\
\end{tabular}}
\end{minipage}
\caption{\label{wine}Wine Data: 10-fold OOS prediction experiment Root Mean Square Error and \% Worse than Best for the mean RMSE. }

\end{figure}

We consider two additional prediction problems.  The first example, 
taken from \cite{CorCer09}, involves prediction of an `expert' quality ranking on
the scale of 0-10 for wine based upon 11 continuous attributes (physiochemical properties of the wine) plus
wine color (red or white).  There are 4898 observations.  Results are in Figure \ref{wine}:  EBF is only 1\% worse than the full BF, while SSF is 12\% worse.

\begin{figure}
\begin{minipage}{0.5\linewidth}
\includegraphics[width=\textwidth]{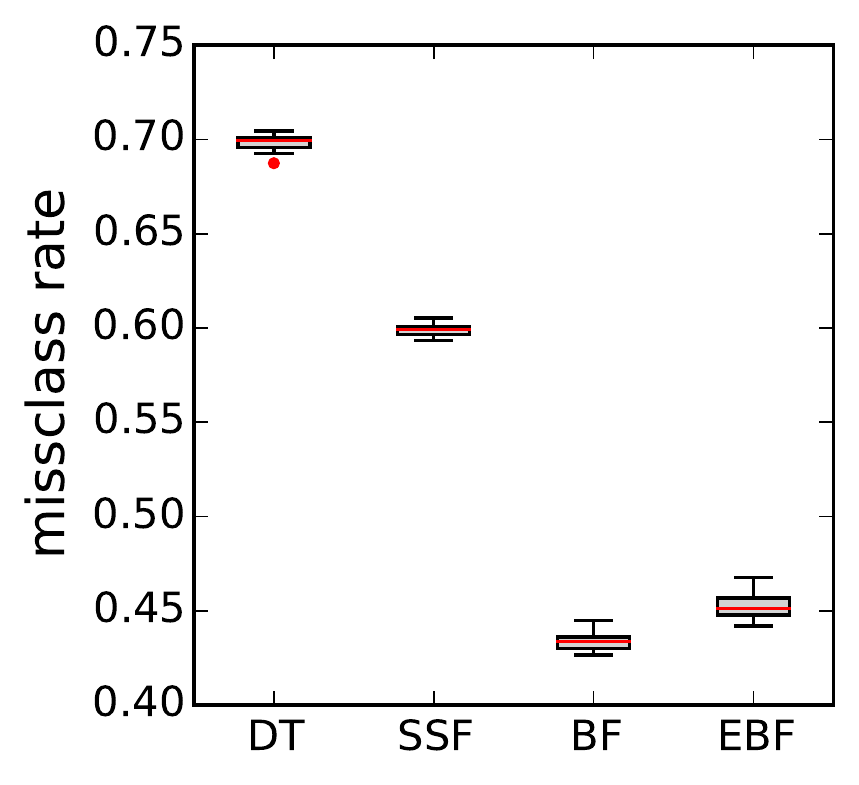}
\end{minipage}
~~
\begin{minipage}{0.4\linewidth}
{\footnotesize
\begin{tabular}{c c | l}
$\overline{\text{MCR}}$  & \% WTB & \\
\cline{1-2}\rule{0pt}{3ex} 
0.4341 &  0.0 & BF \\
0.4531 &  4.4 & EBF \\
0.5989 & 38.0 & SSF \\
0.6979 & 60.8 & DT \\
\end{tabular}}
\end{minipage}
\caption{\label{beer} Beer Data: 10-fold OOS prediction experiment Miss-Classification Rate and \% Worse than Best for the average MCR. }
\end{figure}

The second example is from the Nielson Consumer Panel data, available for
academic research through the Kilts Center at Chicago Booth,
 and our sample contains 73,128 purchases
of light beer in a number of US markets  during 2004. The response of interest
is brand choice, of a possible five major light beer labels: Budweiser,
Miller, Coors, Natural, and Busch. Each purchase is associated with a
household that is codified through 16 standard demographic categorical
variables (maximum age, total income, main occupation, etc).  Applying
classification forests and DTs (based on Gini impurity) leads to the results
in Figure \ref{beer}:  EBF is only 4.4\% worse than the BF, while  SSF is 38\%
worse.

\subsection{Choosing the trunk depth}

From the distributed computing perspective which motivates this work, you should fix the trunk only as deep as you must.  In our full scale applications (next section), this means setting branch size so that the working memory required to fit a forest at each branch is slightly less than the memory available on each machine.

This leaves open questions, e.g., on the trade-off between
number of trees in the forest and depth of the trunk.  We don't yet have
rigorous answers, but the npB framework can help in further
investigation. In the interim, Table \ref{msltab} shows the effect on OOS
error from doubling and halving the minimum leaf size for the EBFs in our
three examples.

\begin{table}[h]
{\footnotesize
\begin{tabular}{c | c c c | c c c | c c c}
& \multicolumn{3}{l|}{CA housing} & \multicolumn{3}{l|}{Wine} &\multicolumn{3}{l}{Beer} \\
\cline{2-10} \rule{0pt}{3ex} 
\!\!\!\!\!\!MLS $10^3$ & 6 & 3 & 1.5 & 2  & 1 & 0.5 & 20 & 10 & 5\\
\!\!\!\!\!\!\% WTB & \!1.6 & \!2.4 & \!4.3 & \!0.3 & \!0.8 & \!2.2 & \!1.0 & \!4.4 & \!7.6 
\end{tabular}}
\caption{\label{msltab}\% Worse than Best on OOS error for different minimum leaf sizes (MLS) specified in thousands of observations.}
\end{table}

\section{Scaling for Big Data}

To close, we note that this work is motivated by the need for reliable forest fitting algorithms that can be deployed on millions or hundreds of millions of observations, as encountered when analyzing internet commerce data.  A number of additional engineering details are required for such deployment, but the basic approach is straightforward:  an initial trunk is fit (possibly to a data subset)\footnote{Note that the original CART trunk can itself be fit in distribution, e.g. using the \texttt{MLLib} library for Apache Spark.}, and this trunk  acts as a sorting function to map observations to 
separate locations corresponding to each branch.  Forests are then fit at each location (machine) for each branch.  

Preliminary work at eBay.com applies EBFs for prediction of `Bad Buyer Experiences' (e.g. complaints, returns, or shipping problems) on the site.  Training on a relatively small sample of 12 million transactions, the EBF algorithm using 32 branch chunks is able to provide a 1.3\% drop in misclassification over the SSF alternatives.  This amounts to more than 20,000 extra detected BBE occurrences over the short sample window, potentially giving eBay the opportunity to try and stop them before they occur.

\section*{Acknowledgments} 
 
Taddy is also a scientific consultant at eBay and a Neubauer
family faculty fellow at the University of Chicago.

\bibliography{forests}
\bibliographystyle{icml2015}

\end{document}